
\magnification=\magstep1
\voffset=-0.8 true in


\newcount\equationno      \equationno=0
\newtoks\chapterno \xdef\chapterno{}
\def\eqn{\eqno\eqname}
\def\eqname#1{\global \advance \equationno by 1 \relax
\xdef#1{{\noexpand{\rm}(\chapterno\number\equationno)}}#1}


\def\ga{\mathrel{\mathchoice {\vcenter{\offinterlineskip\halign{\hfil
$\displaystyle##$\hfil\cr>\cr\sim\cr}}}
{\vcenter{\offinterlineskip\halign{\hfil$\textstyle##$\hfil\cr>\cr\sim\cr}}}
{\vcenter{\offinterlineskip\halign{\hfil$\scriptstyle##$\hfil\cr>\cr\sim\cr}}}
{\vcenter{\offinterlineskip\halign{\hfil$\scriptscriptstyle##$\hfil\cr>\cr\sim\cr}}}}}

\def\hmpc{h^{-1}{\rm Mpc}}
\centerline{\bf CONSTRAINTS ON THE MODELS FOR STRUCTURE FORMATION FROM}
\centerline{\bf  THE ABUNDANCE OF DAMPED LYMAN ALPHA SYSTEMS  }
\vskip 2 true cm
\centerline{\bf K.Subramanian$^{1}$ and T.Padmanabhan$^2$}
\vskip 1 true cm
\centerline{$^1$ National Center for Radio Astrophysics, TIFR}
\centerline{Post Bag 3,Ganeshkhind,Pune 411 007,India}
\smallskip
\centerline{$^2$Inter University Center for Astronomy and Astrophysics,}
\centerline{Post Bag 4,Ganeshkhind,Pune 411 007,India}
\smallskip

\vskip 1 true cm
\centerline{IUCAA-5/94; Jan,94; submitted for publication}
\vskip 2 true cm
\noindent{\bf Models for structure formation attempt to predict the power
spectrum of density
perturbations in the present universe from the initial power spectrum
and the nature of dark matter.
Observational constraints on the power spectrum at different scales
in the present epoch can, therefore, be used to eliminate (or choose between)
different theoretical
models. Such a comparison is fairly easy at large
scales (at which linear theory is valid), and one can use observations like
the MBR anisotropy, large scale steaming motions etc
to constrain the models. But to
discriminate between the models effectivley, it is
necessay to constrain the power spectrum
at small scales.
 The most reliable constraints on the power spectra at small
scales come from the predicted abundance of bound systems which can
be estimated reasonably accurately using Press-Schecter
(or similar) methods$^1$.
In the past, this method has been used in conjunction
with the quasar abundance$^{2-4}$
and cluster abundance$^{5-7}$. We show here that the
abundance of damped lyman alpha
systems (DLAS, hereafter), provides a far stronger constraint on the models
for structure formation. Models with a mixture of
hot and cold dark matter$^{8-11}$ (which are consistent
with large scale observations) are strongly ruled out by the DLAS
constraints
while models with cosmological constant$^{12}$ are marginally inconsistent.
It is also possible to combine the constraints from the abundance of clusters,
DLAS and QSO's to obtain model-independent bounds
on the power spectrum at the nonlinear scales. These bounds are to be respected
by any viable model
for structure formation.}

The damped Lyman $\alpha$ systems have been studied extensively in the recent
years. The spectroscopic surveys for DLAS have shown that$^{13-16}$
the absorbers - found between redshifts $1.7$ to $4$ - have an
average HI column density $<\bar N> \sim 10^{21} {\rm cm}^{-2} $.
The most striking property of the DLAS comes from a determination
 of their abundance characterised by the
mean number density of lines per unit redshift, $dN/dz$.
Lanzetta et al (1991) find$^{16}$ the number of damped
Ly$\alpha$ systems per unit redshift interval with
$\bar N \ga 2\times 10^{20} {\rm cm}^{-2} $ is
$$ 0.16 \pm 0.03 < {dN_{damp}\over dz} < 0.25 \pm 0.04  \qquad {\rm at} \quad
<z> = 2.5 .\eqn\ndamp $$
Here $<z>$ refers to the average redshift of these systems and the
two bounds are
derived from two distinct subsamples of the data.
Their data is also consistent with
there being no evolution in the absorber properties.
The number density of the damped systems
exceeds that expected from present day galaxies by a factor of about $3-5$
(see ref. 16).
This has been interpreted to indicate that the cross section
of the absorbers is larger than that associated with galaxies.

More importantly, the observed value of
$dN_{damp}/dz$, used in conjunction with the mean column density
allows one to estimate the density of neutral hydrogen contributed by the DLAS.
The mean mass density of neutral gas contributed by the damped Ly$\alpha$
systems
at their mean redhift $<z>$ is
$\rho_{damp}(<z>)= \mu m_p < \bar N > (dN /cdt)$.
Here $\mu =1.4$ is the mean molecular weight of the gas and $m_p$ is
the proton mass. As the universe expands this average
mass density would have decreased by a factor $(1 + <z>)^3$. Comparing the
resulting density with the present day critical density one gets
the current density parameter of the HI making up the DLAS
to be
$$\Omega_{damp}=\mu m_p <\bar N> {dN\over dz}\vert_{<z>}
{(1 + 2q_0<z>)^{1/2}\over \rho_c (1 +<z>)}\left({H_0\over c}\right) \eqn\omdam
$$
(This result can be easily be modified
to take into account a cosmological constant).
To obtain the total mass associated with the DLAS we have
to include two factors:
Suppose $f_N$ is the HI fraction of the gas in the absorbers and
$f_b$ is the baryonic fraction in the universe. Then the density parameter
contributed by
the total mass associated with the DLAS,
which has collapsed to form bound objects is
$$ \Omega_{DLAS} = {\Omega_{damp}\over (f_N f_b) } \approx
{0.352 \over (1 +z)^{1/2}} h^{-1} \left({dN/dz \over 0.2}\right)
\left({<\bar N> \over 10^{21}
cm^{-2}}\right)\left({f_N\over 0.5}\right)^{-1} \left({f_b\over 0.015
h^{-2}}\right)^{-1} \eqn\fcol$$
for a $\Omega =1$ universe.
At an average redshift of $2.5$ one gets $\Omega_{DLAS} = 0.094$ if
$h=0.5$.

It is not easy to estimate the mass associated with the individual DLAS
directly. However, there are several indications to suggest that
these systems have masses in the range of $10^{11-12}M_\odot$.
If we make this assumption that DLAS are galactic scale objects,
then  we see from \fcol\ that their abundance will provide
the most stringent constraint on models of structure formation :
any such model should have enough power on galactic
scales to be consistent with a collapsed fraction of about $10 $ percent, at
the redshifts associated with DLAS.
We elaborate on this point in
greater detail below after considering briefly the evidence that the
damped systems indeed represent galactic mass objects.

For $dN/dz=0.2$, a typical value indicated by \ndamp\ , one
infers $\Omega_{damp} = 1.5h^{-1} \times 10^{-3}$ for a flat universe.
By comparison the mass density of the
stars in present day galaxies are thought$^{17}$ to account for
$\Omega_{vis} \sim 3 h^{-1} \times 10^{-3}$.
This similarity of $\Omega_{damp}$ and $\Omega_{vis}$ first led to the
suggestion that the damped Ly$\alpha$ systems may be the progenitors of
present day luminous galaxies observed before the bulk of their stars
formed, when the gas fraction was high.
An alternate interpretation$^{18}$ is that these systems may represent a
population
of gas rich dwarf galaxies, which had smaller cross section but were much
more abundant in the past.
Direct information on the size
and mass of the damped Ly$\alpha$ absorbers is limited to only a few
cases at present: Study of the 21 cm absorption line$^{19}$ caused by the
$z = 2.04$ DLAS in
PKS 0458-02, indicates that the absorber must
extend more than $8 h^{-1} {\rm kpc}$ across the line of sight. Spatial
imaging$^{20}$ of the Ly$\alpha$ emission at $z=2.811$ towards the
QSO 0528-250, shows that the DLAS at $z=2.811$ could be a compact group of
gas rich galaxies or a single large cloud of diameter
$\sim 100$ kpc. Detection of CO emission lines$^{21}$
from the $z=2.14$ DLAS in the same QSO indicates that the absorber may be a
group of galaxies each galaxy having a gas mass
$\sim 2\times 10^{11} M_\odot$ and virial dimension $\sim 40$ kpc.
Deep imaging of the field around the QSO 0000-263 has revealed a candidate
galaxy associated with the $z=3.34$ DLAS seen along the line of sight.$^{22}$.
In view of these observations, it appears likely
that the damped Ly$\alpha$ systems
are gas rich massive galaxies rather than dwarfs. This is also
consistent with the finding$^{23}$ that luminous galaxies, rather than dwarfs,
are
responsible for producing the MgII absorption systems, of which the damped
systems are a subset, at $z= 0.3 -0.9$. We shall explore
the consequences of this point of view below.

To do this we have to compute the theoretically expected value of
 $\Omega_{theory}(M,z)$ contributed by bound systems with mass
greater than $M$ at any given redshift $z$. This can be done using the
Press-Schecter
formalism and we get the result that:
$$\Omega_{theory}(>M,z) = erfc\left[ {\delta_c(1+z)^2\over
\sqrt{2} \sigma_0(M)}\right] \eqn\ff$$
where erfc denotes the complementary error function, $\sigma_0(M)$ is
the power spectrum evaluated at $z=0$ by the linear theory and $\delta_c$ is
the linearly extrapolated density contrast needed for collapse.
In the simple spherical top hat model one finds that $\delta_c=1.68$.
Given a theoretical model, we know $\sigma_0(M)$ and hence can estimate
$\Omega_{theory}$. For a model to be viable we need $\Omega_{DLAS}\leq
\Omega_{theory}$, with the equality implying that every collapsed object
is hosting a DLA system. This restriction, it turns out, can constrain the
models very effectively.

In figure 1 we have plotted $\Omega_{theory}(10^{12}M_\odot,z)$,
the fraction of the mass
which has collapsed into objects with
$ 10^{12} M_\odot< M < 10^{13} M_\odot $, for different theoretical models
and compared them with the $\Omega_{DLAS}(z)$ inferred from
the data of Lanzetta et al.$^{16}$, using \fcol\ .  The data points of
Lanzetta et al. have been given by Turner and Ikuechi$^{24}$ as a table
of $dN/dz$ at different redshifts. The data points with the error bars are for
an $\Omega=1$ universe. If $\Omega_{matter}<1$, due to the presence of a
non-zero
cosmological constant, then the data needs to be shifted up. For
$\Omega_{vac}=0.8$,
the centers of the data need to be shifted to the location marked by the
crosses.

 All the theoretical models in figure 1 are based on a primordial
power spectrum of the form $P(k)\propto k$.
The solid curve is for the pure CDM model with the BBKS transfer function; the
broken curve
is for a model with CDM and cosmological constant (called V+CDM model herafter)
 with $\Omega_{vac}=0.8,\Omega_{cdm}=0.2$;
the dotted curve is for the CDM model with a bais factor of 2.5 (such a model,
of course,
cannot reproduce large scale observations including the COBE results unless the
bias factor is scale dependent; we have included it only for the sake of
comparison). The dot-dash curve is for a mixed H+CDM model with $\Omega_{hdm}=
0.7,\Omega_{cdm}=0.3$. All models (except, of course, the b=2.5 one) are
normalised
 so as
to reproduce the MBR quadrapole anisotropy measured by COBE. We have also taken
h=0.5 for all cases except for the V+CDM model for which h=0.8. These
parameters are
chosen so that the models are consistent with the large scale observations.

The comparison shows that the mixed dark matter models are strongly ruled out
by
the abundance of DLAS. The disparity is so high that the theoretical and
observational
uncertainties in the analysis is unlikely to be of any use in saving the model.
The
V+CDM model also falls short of producing the required abundance (note that in
this case
the data points are shifted up; it is the location of the crosses which one has
to
compare with the theoretical curves). However, the uncertainties in the data as
well
as some freedom which is available in theoretical modelling can be utilised to
make the disparity lower. We feel that the data does rule this model also out
though
not as firmly and conclusively the H+CDM model. The pure CDM with COBE
normalisation,
of course, has no difficulty in explaining the abundances since it has a lot of
small
scale power; however this model is known to be inconsistent with other
observations
(like results from galaxy surveys). Finally, the (once popular) CDM model with
b=2.5
also falls short of explaining the DLAS abundances.

In the past, these models were tested as regards the observed quasar abundance.
The constraints posed by the abundance of DLAS are far more stringent. To see
this,
let us calculate the $\Omega_{quasar}$, the density contributed by the mass
associated with the quasars, along the folowing lines:
 The basic strategy is to
relate the observed quasar abundance to the abundance of host galaxies
of quasars. Quasars are thought to be powered by accretion on to black holes.
Unless special models using super Eddington luminosities are involved, one can
infer a characteristic black hole mass by assuming that
the observed luminosity of the quasar corresponds
to the Eddington luminosity.
The mass of the host galaxy associated
with a quasar of blue magnitude $M_B$ can then be estimated to be:
$$M_G \approx  c_1 \times 10^{12}M_\odot \times
10^{0.4[-M_B -26]} \eqn\mg $$
where
$$c_1 = 1.2 ({f_{bol}\over 10})({f_b \over 0.06})^{-1}
({f_{hole} \over 0.01})^{-1} .\eqn\ci$$
The three f-factors arise as follows: To convert the blue magnitude to
the bolometric magnitude we have to use a correction factor, $f_{bol}$, which
depends
on the quasar spectrum
but is in general
of order of about $10$. Secondly, only a fraction
$f_b$ of matter in the universe may be in baryonic form.
And, finally, only a fraction $f_{hole}$ of the baryons in a collapsing host
galaxy
be able to form the compact central object.
We find from \mg\  that a bright quasar with $M_B = -26$  is typically
hosted by a galaxy of mass of about $ 10^{12} M_\odot$.

The density contributed by the quasars can now be written as
$$\Omega_{quasar}(>M,z) = \rho^{-1} \int_{-\infty}^{M_B(M)}
\tau M_G(M_B) \Phi(M_B,z) dM_B \eqn\freq$$
where $M_B(M)$ is got by inverting \mg\ , $\Phi$ is the quasar
luminosity function  and $\rho$ is the mass density of the
universe, at present. Further $\tau = max (1, t(z)/t_Q)$,
takes into account the fact
that only a fraction of order $t_Q/t(z)$ of galaxies will display quasar
activity if the quasar lifetime is smaller than
than the age of the universe $t(z)$ at redshift $z$.
For the quasar luminosity function we shall use the form
 derived by Boyle et al.$^{25}$
for a $\Omega=1$, $h=0.5$ universe. We shall also
assume that their luminosity function, which is unevolving beyond $z=1.9$,
can be used upto $z=4$ (see Ref. 4).
Consider the abundance of collapsed halos required to explain the
most luminous quasars, with $M_B < -26$.
One can evaluate the integral in \freq\ numerically to find
$$\Omega_{quasar}(> c_1 10^{12} M_\odot,z > 1.9)= 1.4 c_1 \times 10^{-3}
\eqn\fneed$$
For a particular model of structutre formation to successfully explain the
abundance of quasars it must give $\Omega_{theory}(>M,z) >
\Omega_{quasar}(>M,z)$ .
In figure 1 we have also plotted this line for $1.9 < z < 4.0$. While this also
rules out H+CDM model, it is clear that the constraint arising from this
consideration
is weaker than the one from DLAS.

The abundance of objects like DLAS, quasars (and clusters) can be used to
obtain
bounds on the power spectrum $\sigma_0(M)$ at different scales. This is shown
in
figure 2 which is a plot of $\Omega$ versus $\sigma$ for different values of z.
This plot is clearly independent of the power spectrum in the model and depends
only on the validity of Press-Schecter analysis. The observed abundances of
different objects at different
redshifts can be used to mark out regions in this graph, thereby
constraining the
value of $\sigma$ at the mass scales corresponding to the objects. For example,
the abundunce of Abell clusters at $z\simeq 0$ gives
$\Omega\simeq(0.001-0.02)$;
from the plot (marked by short horizontal lines in the $z=0$ curve)
 we see that this leads to the tight constraint of $\sigma\simeq(0.5-0.7)$
at cluster scales [which turns out to be around $ 8\hmpc$]. Similarly, the
quasar
abundance (marked by a straight horizontal line) gives $\sigma>(2-2.5)$ at
galactic scales.
 The DLAS give $\sigma>(3-4)$
at the same scale (see the shaded region). Any model for structure formation
should
provide sufficient power at these
galactic and cluster scales in order to be compatible with the observations.

Similar constraints on the models for structure formation (from the abundance
of DLAS)
 has been reached independently in ref. 26.

\beginsection Aknowledgements

One of the authors (T.P) would like to thank M.J. Rees, J.Miralda-Escude and
H.Mo
for discussions
and the Institute of Astronomy, Cambridge for hospitality during a visit by T.P
in
summer,93.

\beginsection{\bf References}

\item{1.} Press, W.H. $\&$ Schecter, P.L., Ap.J., {\bf 187},425   (1974).
\item{2.} Efstathiou, G. $\&$ Rees, M.J., M.N.R.A.S., {\bf 230}, 5p  (1988).
\item{3.} Kashlinsky, A. $\&$ Jones, B.J.T., Nature, {\bf 349}, 753  (1991).
\item{4.} Haehnelt, M.G., M.N.R.A.S.,{\bf 265}, 727 (1993).
\item{5.} White,S.D.M, G.Efstathiou and C.S.Frenk, M.N.R.A.S, {\bf 262},
1023-1028 (1993).
\item{6.} Bond,J.R, Myers, S.T in {\it Trends in Astroparticle physics},
eds., D.Cline and R.Peccei, (World Scientific,Singapore, 1992) p 262.
\item{7.} Lilje,P.B, Astrophys. J., {\bf 386},L33.
\item{8.} Taylor, A.N. $\&$ Rowan-Robinson, M., Nature, {\bf 359},396 (1992).
\item{9.} Davis. M. et al., Nature, {\bf 359}, 393 (1992).
\item{10.} Klypin et al., Ap. J.,{\bf 416},1 (1993).
\item{11.} Pogosyan, D.Y. $\&$ Starobinsky, A.A., M.N.R.A.S, {\bf
265},507(1993)
\item{12.} Kofman, L. et al., Astrophys. J., {\bf 413}, 1(1993)
\item{13.} Wolfe, A.M., Turnshek, D.A., Smith, H.E. $\&$ Cohen, R.D.,
Astrophys. J. Suppl., {\bf 61}, 249 - 304 (1986).
\item{14.} Wolfe, A.M. {\it The Epoch of Galaxy Formation}
(eds. Frenk, C.S., Ellis, R.S., Shanks, T., Heavens, A.F., Peacock, J.A. )
101 - 105 (Kluwer, Dordrecht, 1989).
\item{15.} Turnshek, D.A., Wolfe, A.M., Lanzetta, K.M., Briggs, F.H.,
Cohen, R.D., Foltz, C.B., Smith, H.E. $\&$ Wilkes, B.J.,
Astrophys. J., {\bf 344}, 567 - 596 (1989).
\item{16.} Lanzetta, K.M., Wolfe, A.M., Turnshek, D.A., Lu, L.,
McMahon, R.G. $\&$ Hazard, C., Astrophys. J. Suppl., {\bf 77}, 1 - 57 (1991).
\item{17.} Gnedin, N., Ostriker, J.P., Ap. J., {\bf 400}, 1 (1992).
\item{18.} Tyson, N.D., Ap. J. Lett., {\bf 329}, L57  (1988).
\item{19.} Briggs, F.H. et al., Ap. J., {\bf 341}, 650  (1989).
\item{20.} Moller, P. $\&$ Warren, S.J., Astron, Ap. (in Press) (1993).
\item{21.} Brown, R.L. $\&$ Vanden Bout, P.A., Ap. J. Lett., {\bf 412}, L21,
(1993).
\item{22.} Steidel, C.C $\&$ Hamilton,D, Astrophys.J, {\bf 104}, 941 (1992).
\item{23.} Bergeron, J. $\&$ Boisse, P., A $\&$ A, {\bf 243}, 344 (1991).
\item{24.} Turner, E., Ikeuchi, S., Ap. J., {\bf 389}, 478 (1993).
\item{25.} Boyle, B.J. et al., in {\it ASP conference series No. 21. The
space distribution of quasars.}, ed. Crampton, D. (ASP, San Fransisco), 191-
(1991).
\item{26.} Mo,H and J.Miralda-Escude,1994, preprint; submitted to Ap.J.Letts.
\beginsection Figure Captions

Figure 1: The density contributed by collapsed objects with mass in the
range of $10^{12}-10^{13}$ in various theoretical models is compared with
observations. All the theoretical models are based on a primordial
power spectrum of the form $P(k)\propto k$.
The solid curve is for the pure CDM model with the BBKS transfer function; the
broken curve
is for a model with CDM and cosmological constant (called V+CDM model herafter)
 with $\Omega_{vac}=0.8,\Omega_{cdm}=0.2$;
the dotted curve is for the CDM model with a bais factor of 2.5.
 The dot-dash curve is for a mixed H+CDM model with $\Omega_{hdm}=
0.7,\Omega_{cdm}=0.3$. All models (except, of course, the b=2.5 one) are
normalised
 so as
to reproduce the MBR quadrapole anisotropy measured by COBE. We have also taken
h=0.5 for all cases except for the V+CDM model for which h=0.8. The data points
 with the error bars are based on the observed abundance of DLAS and are for
an $\Omega=1$ universe. If $\Omega_{matter}<1$, due to the presence of a
non-zero
cosmological constant, then the data needs to be shifted up. For
$\Omega_{vac}=0.8$,
the centers of the data need to be shifted to the location marked by the
crosses.
The
horizontal line is based on the abundance of quasars.

\noindent
Figure 2: The $\Omega$ contributed by collapsed objects
 is plotted against the linearly extrapolated
density contrast $\sigma_0 $. The curves are parmetrised by the redshifts
$z=0,1,2,3,4$
from top to bottom. The observed abundances of
different objects at different
redshifts are used to mark out regions in this graph, thereby
constraining the
value of $\sigma$ at the mass scales corresponding to the objects.
The abundunce of Abell clusters at $z\simeq 0$ (marked by short horizontal
lines in the $z=0$ curve)
 gives $\Omega\simeq(0.001-0.02)$ and
 leads to the tight constraint of $\sigma\simeq(0.5-0.7)$
at cluster scales [which turns out to be around $ 8\hmpc$]. Similarly, the
quasar
abundance (marked by a straight horizontal line) gives $\sigma>(2-2.5)$ at
galactic scales.
 The DLAS give $\sigma>(3-4)$
at the same scale (see the shaded region) thereby providing the most stringent
constraint.

\end